\documentclass{article}
\usepackage{graphicx}
\usepackage{amsmath}
\usepackage{amsthm} 
\usepackage{amssymb}
\usepackage{color}
\usepackage{textcomp}
\usepackage{rotating}
\usepackage{pdfpages}

\begin{document}

\title{The ASDC SED Builder Tool description and Tutorial}
\author{G. Stratta, M. Capalbi, P. Giommi, R. Primavera, S. Cutini, D. Gasparrini \\
on behalf of the ASDC team}

\maketitle

\abstract

The ASDC SED Builder (http://tools.asdc.asi.it/SED/) is a web based program developed at the ASI Science Data Center to produce and display the Spectral Energy Distribution (SED) of astrophysical sources. The tool combines data from several missions and experiments, both ground and space-based, together with catalogs and archival data. In the current version (v1.3) the obtained SEDs can be compared with theoretical expectations and with the sensitivity curve of several widely known instruments. The displayed data can also be fitted to simple analytical functions. Providing a cosmological redshift, the SED can be visualized in rest-frame luminosities. The tool provides transparent access to ASDC-resident catalogs (e.g. Swift, AGILE, Fermi etc.) as well as to external archives (e.g. NED, 2MASS, SDSS etc.) covering the whole electromagnetic spectrum, from radio to TeV energies. Proprietary data can also be properly handled. The intent of this document is to provide a brief description of the main capabilities of the ASDC SED Builder. Specific details on the graphical interface and on the functionalities can be found in the appendix to this document which provides a tutorial\footnote{http:$//$www.asdc.asi.it$/$tutorial$/$SEDBuilder$/$SEDBuilderTutorial.html}  to the tool.

\section*{Introduction}

Spectral Energy Distributions (SEDs) are commonly used in contemporary multi-wavelength  astronomy because they are considered to be a powerful instrument to study the physical properties of astrophysical sources by providing a direct comparison of source luminosities in different energy bands.

The ASDC SED Builder tool has been developed at the ASI Science Data Center\footnote{http:$//$www.asdc.asi.it} (ASDC) to allow web users to  easily browse and combine multi-wavelength data to produce a SED of any cosmic source. There are several catalogs in the ASDC databases, which are periodically updated; transparent queries to external services can also be performed. 

In particular, many high energy astronomy catalogs are available (e.g. Fermi, AGILE, Swift XRT). For these catalogs data are often not available in physical units but rather in count rates measured by specific instruments. In these cases, data are converted to fluxes taking into account the instrument response and assuming a spectral model. This operation is done for each catalog containing count rates. The correction for the Galactic absorption is also applied.

The tool, based on a Java code and a MySQL database system, provides different functionalities and several plot options for the analysis of the SEDs. The possibility to filter the data in time intervals will be added to the next version, thus enabling the construction of time resolved SEDs suitable for variability studies.

The need for a SED builder tool has been listed among the  priorities of the 
International Virtual Observatory Alliance (IVOA). For this reason some groups are planning specific tools for the near future (e.g. D'Abrusco et al. 2010, arXiv:1012.5733). On our side, we plan to make the ASDC SED Builder VO compliant within a few months, starting by adding the option to save the SED data as VO table and to query the VO registry.

\section*{Building SEDs}

Given a selected astrophysical source, flux information at several wavelengths necessary for the SED building are gathered browsing through  several missions and experiments databases, both ground and space-based.
To achieve this goal, the ASDC SED Builder welcome page (http://tools.asdc.asi.it/SED/) redirects the user to a dedicated query page from which the source name or coordinates can be specified. Some specific information for that object such as the name and sky coordinates will be displayed at the bottom of the page. Note that the result of a search through sky coordinates depends strongly on the search radius (1 arcmin by default) and it can produce more than one entry. 

All catalogs containing the queried coordinates are loaded by default for the SED building (see an example in Figure 1 for the bright quasar 3C279). 
The user can refine the input data by selecting/deselecting different catalogs, by changing the search radius of each catalog and by using several plotting options. 

\begin{figure}
\centering
\includegraphics[width=10cm,angle=0]{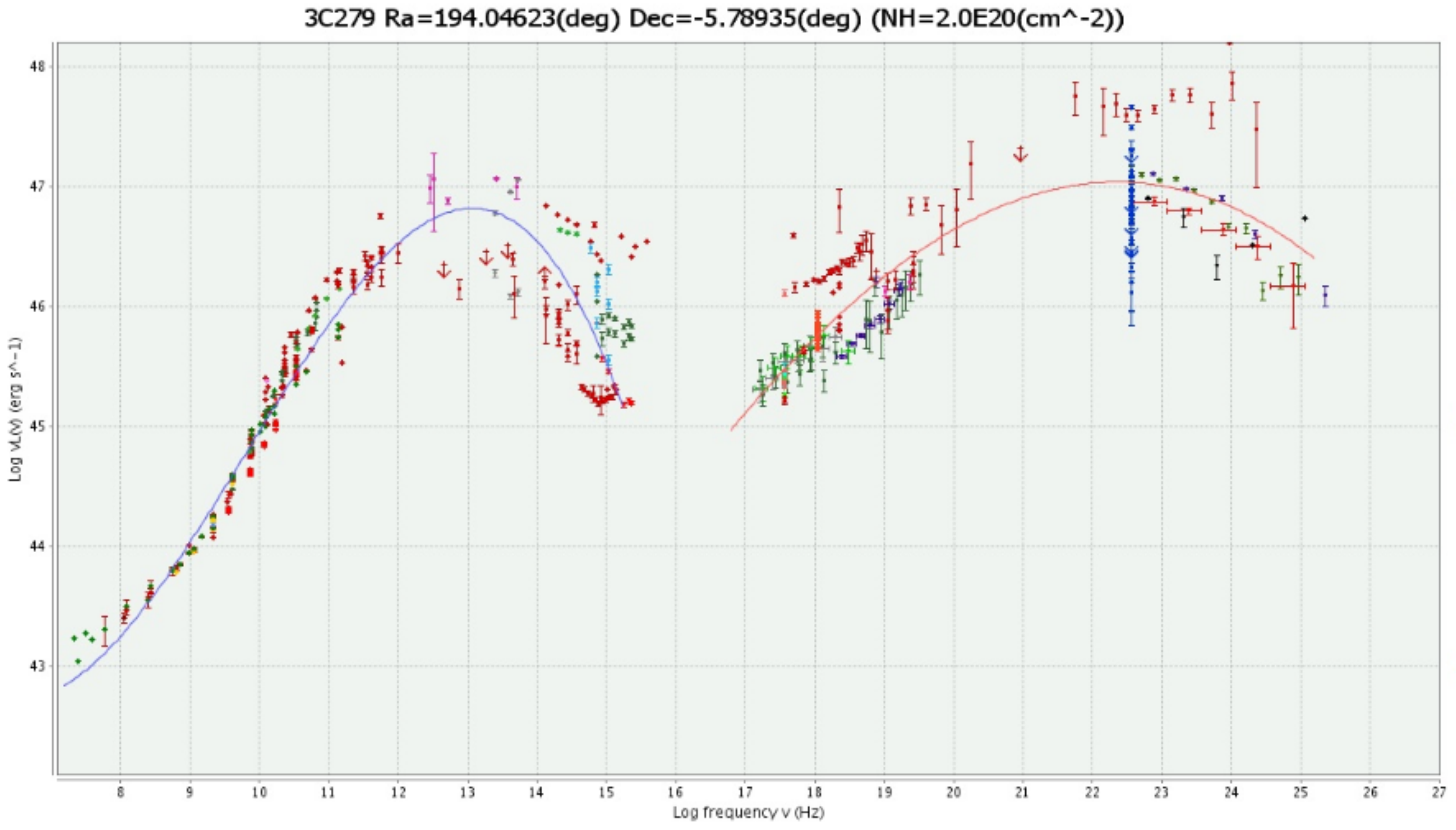}
\caption{The SED of 3C279 obtained with the ASDC SED Builder tool.\label{fig1}}
\end{figure}

\section*{SED Input Data}

Input data are divided in three different categories:

\begin{itemize}

\item{{\bf Local Catalogs}}
 
The ASDC Team supplies the user with a list of local catalogs grouped by frequency ranges: radio, infrared, optical, X-ray and gamma-ray

\item{{\bf External Catalogs}}
 
Data provided by external multi-wavelength databases/surveys, such as NED, 2Mass, USNO, and SDSS

\item{\bf User Catalogs}

Proprietary data/catalogs, handled in a dedicated area, can be uploaded
\end{itemize}

A dedicated ``User Data" area enables to upload and manage in a secure way proprietary (or any preferred) data. The user can upload single source or multiple sources new data using a simple ASCII file and following a simple syntax. 
After uploading and saving, the new set of data will be available as ``User Catalogs". Proprietary data can be managed only by registered users and by default are not accessible to other users: however it is possible to share data with other groups and/or single users. 

All loaded data can be visualized through an interactive table. All sources located within the circular region defined by the search radius of each catalog are listed in terms of sky coordinates, flux density and frequency. Each SED point can be selected/deselected from each table and the new configuration can be updated (e.g. to exclude a particular energy range). This operation is particularly useful to clean the obtained SED from spurious entries that may have been found in the automatic search procedure. 
Once cleaned, the built SED can be analyzed by the user in several ways directly from the tool.

\section*{SED Analysis}

A number of functionalities are currently implemented in the ASDC SED Builder web tool to enable the user to manage input data and compare them to theoretical expectations or to the sensitivity curves of several instruments, by performing fit with simple analytical functions, by tailoring the plot settings, etc. 

In the present version of ASDC SED Builder, two types of Synchrotron Self Compton emission model are available:
\begin{itemize}
\item {\bf SSC Analytical} provides an analytic solution of the emission model obtained through a number of approximations (developed by A. Pellizzoni and M. Perri). The model is based on a broken power law electron distribution.

\item{\bf SSC Numerical} provides a more precise but somehow slower numerical approach to the model. It allows the user to choose among different electron distributions (e.g. power law, broken power law, log parabola, etc.). The user can select the model accuracy level and whether to consider or not the Synchrotron self-absorption component. Publications using this code should cite the following paper: A. Tramacere et al 2009, A\&A, 501, 879.

\end{itemize}

Both Analytical and Numerical SSC models require the knowledge of the source redshift. The user can plot both models together over the data and sum existing single models creating a model grouping. 

Two useful spectral templates are also provided: a composite QSO optical spectrum from SDSS (D.E. Vanden Berk, 2001, ApJ, 122,549) plus standard X-ray emission (radio to optical flux ratio is required) and a giant elliptical galaxy template (derived from Mannucci et al. 2001, MNRAS 326, 745). Polynomial functions can be fitted to the built SED, in any desired frequency range: the main goal of this functionality is to enable the user to estimate any peak flux value in the displayed SED, quoted as ''Max value" at the bottom of the frame as soon as the fit has been performed. 

The built SED can also be compared with the maximum sensitivity of several widely known instruments. By selecting each instrument, a sensitivity curve is displayed in the SED plot. We warn the user that an updating work of some instrument sensitivity curve is still in progress.

Registered users can save all built SEDs and SED analysis sessions in the ''User data" area. The user can access directly to saved SEDs from the starting page of the ASDC SED Builder web tool. These SEDs and their analysis can be shared with other collaborators: this can be done by editing the SEDs in the ''User Data" area and selecting the desired group/users to allow the access, however proprietary data are modifiable only by the first owner.

Finally, the built SEDs can be exported as ASCII file (QDP syntax) or as an image in Portable Network Graphic format. The user can play on a large number of settings (e.g. background color, dimension and color of each SED point, logarithmic or linear axis, observer or rest frame at a given redshift, etc.)

The next versions of the ASDC SED Builder web tool will enable time dependent SED analysis and will be improved in a number of functionalities focused towards a complete VO compliant version. 

\newpage

\section*{Appendix}



\section*{Supplementary material}

# Appendix

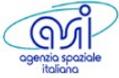 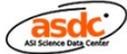 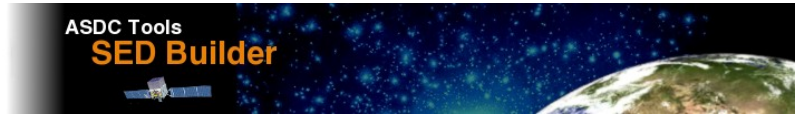

# Tutorial (v1.3.23)

## 1. Introduction

The "ASDC SED Builder" is an advanced tool recently developed at the ASI Science Data Center to allow Web users to manage and display the Spectral Energy Distribution (SED) of astrophysical sources.  The current version has been completely revised in JAVA.
The "ASDC  SED Builder"  combines data from several missions and
experiments, both ground and space-based, together with catalogs and archival data.
The extracted SED can be compared with some theoretical model expectations
and with the sensitivity curve of several popular instruments. The displayed data can also be fitted to simple analytical functions. Finally, providing a cosmological redshift, the frequency range can be shifted to the source rest frame and fluxes can be converted to luminosities.

The "ASDC SED Builder" provides transparent access to all the ASDC public catalogs (e.g. Swift, AGILE, Fermi etc.) as well as external archives (e.g. NED, 2MASS, Sloan etc.), thus covering the whole electromagnetic spectrum,  from radio to TeV energies. Registered users, can use the "ASDC SED Builder" to handle also proprietary data (e.g. user's own data in the appropriate format).

This tutorial provides a guide for the "ASDC SED Builder" graphical interface and an explanation of all the available functions.

## 1.1 SetUp

Version 1.3 of the "ASDC SED Builder"  is available at the
link: http://tools.asdc.asi.it/SED/
The SED Builder allows managing both public and proprietary data.
*Public Data*: without any Login the user will be able to look for existing SED, customize a new one by selecting/deselecting public catalogs, compare data with theoretical expectations, plot available public data, print out SED or download the plot as an ASCII file with QDP syntax.
*Proprietary Data*: registered users only, can use the "ASDC SED Builder" to handle also own data in the appropriate format. Each session can be saved by registered users and potentially shared with other people having the correct data rights.

To register, link to:
https://tools.asdc.asi.it/UserManager/requestUser.jsp
and fill in the form.

# 2. How to build a Spectral Energy Distribution

The "ASDC SED Builder" welcome page (http://tools.asdc.asi.it/SED/) redirects the user to a dedicated query page, from which the source for which we want to build the SED can be specified (Figure 1). The source search can be done in 2 different ways:

<u>Search by Name:</u> From the query page, type the object name then click on "Search by name". All the relevant information for that object will be displayed at the bottom of the page (Figure1). While typing, the system will propose you a list of possible candidates matching your source name: by clicking on the found source from the list the system provides the fill-in boxes with Name and Coordinates.

<u>Search by Coordinates:</u> type the source coordinates and the search radius (default is 1 arcmin), then click on "Search by Coordinates" and all the relevant information for that object will be displayed at the bottom of the page. Note that this search type can produce more than one entries.

From the query output table at the bottom of the page (Figure 1) , by clicking on "Create SED" in the "Tools" column, the user will be prompted to the main page of the SED Builder tool (Figure 2). For registered users, clicking on "SED" at the top right of the page, enables the user to access saved SED, if any.

**Figure 1**  The initial query page where the source for which we want to build a SED can be identified. At the bottom of the frame, there is the output table of the query. Clicking on "Create SED" in the last column will open a new tab the main "ASDC SED Builder" page where the SED can be plotted and managed.

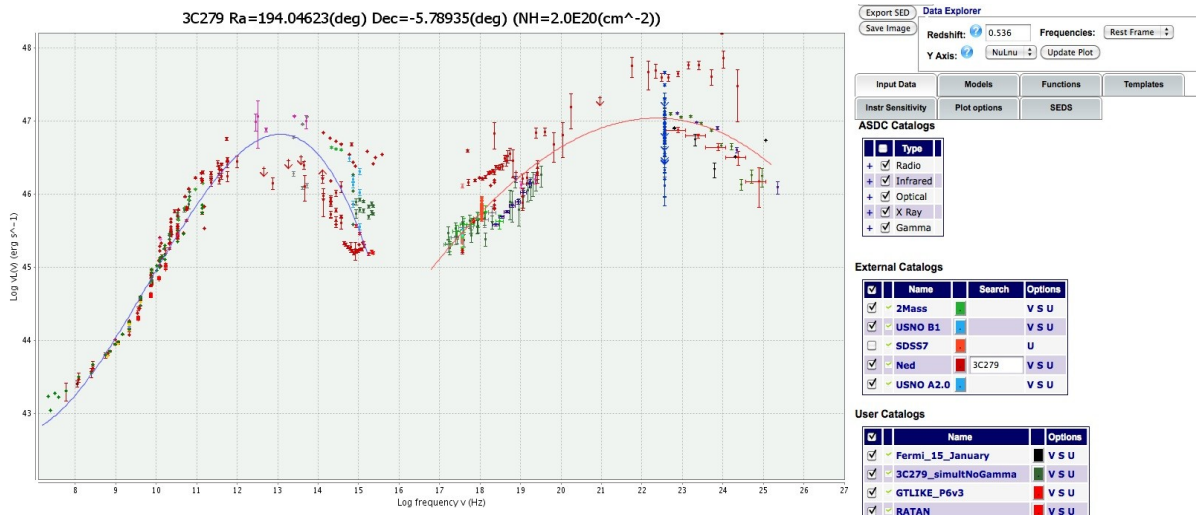

**Figure 2** The "ASDC SED Builder" page.

## 3. The "ASDC SED Builder" Action Buttons

The "ASDC SED Builder" page (Figure 2) contains the SED plot and an operational area on the right side, with several action buttons the function of which is explained below. ***By default, no SED is plotted if "Load Data" button is not clicked.***

**Load Data** enables to plot the SED. By default, all catalogs which contain the queried coordinates are loaded (see section 4.1 for more details on catalog managing). A legend at the bottom of the SED plot shows all the selected catalogs where the queried source has been found. Each catalogue is identified in the plot by a colour defined in the Catalog Tables displayed on the right side (Figure 2). Colours can be modified by clicking on the coloured box next to the "Name" column (but see more possibilities in Section 4.2).

**Show data** opens a new tab which contains the loaded data. For each selected catalog, all entries found in the circular region defined by the search radius of that catalog are listed and their sky coordinates and fluxes at a given frequency are quoted. Note that other sources can be found in the search procedure and the user can exclude them. Indeed, from this window the user can select/deselect the points to be plotted. The new SED can be viewed by clicking "Update". If no modifications have been done, click "Back to SED".

**Export SED** enables the user to save the plotted SED as an ASCII file in QDP format. Different plotting settings can be settled with the "Plot options" dedicated functionality (e.g. linear or logarithmic scale, etc., see Section 4.2 for more details).

**Data Explorer** jump to the "Data Explorer" tool of ASDC that will be opened in a new tab. The "Data Explorer" tool will be automatically uploaded with the coordinates given for the SED (see the "Data Explorer" documentation at http://www.asdc.asi.it/tutorial/DataExplorer/DataExplorerTutorial.html).

**Current SED** this action button is always present during a working session on the top right and enables the user to go back to the SED main page with the last SED plotted.

**Query new SED** enables the user to ask for a new source SED

For registered users only, the following additional Action Buttons will be available:

**User Data** redirects to the user's own Data Area (see more details on this area in Section 5).

**Own SEDs** redirects to your own saved SEDs (see more details in Section 5).

**Save** enables to save the plotted SED in the "Own SEDs" area (see more details in Section 5).

**Edit SED** enables to visualize some basic information on the SED and to modify the SED name and Group Name as well to establish which group is allowed to view the SED (see more details in Section 5.1 and 5.2).

**Duplicate SED** enables to create a copy of the plotted SED in the "Own SED" area.

# 4. The SED Builder dedicated functionalities

A number of dedicated functionalities have been implemented to enable the user to manage input data, to compare data with the theoretical expectations and with the sensitivity curves of several popular instruments, to perform fit with simple analytical functions and to tailor the plot settings. A simple grey-coloured tab set enables the user to easily access to each functionality (see e.g. Figure 2 or Box 1). Each tab content is described below.

## 4.1 Input Data

This functionality enables the user to manage the input data by selecting/deselecting different catalogs, by refining the search radius of each catalog and by playing with some plotting options (see Box 1). The search radius of each catalog can be modified by clicking on the catalog name. At any time, any catalog can be deselected by the user.
The input data are divided in three groups:

> **Local Catalogs** The ASDC Team supplies the user with a list of archival local catalogs arranged by frequency ranges: Radio, Infrared, X-ray and Gamma-ray. Clicking on the "+", catalogues of that frequency range will be listed (see Box 1 for search region and plotting options).

**External Catalogs** Data from external multi-wavelength databases/projects, such as NED, 2Mass and USNO can be uploaded. Note that NED query requests the Object Name.

For registered users only:

**Users Catalogues** shows user's own catalogs of that specific source. Own new catalogues can be uploaded through the **User Data** area (see Section 5).

By clicking on the **Name** of each catalog, it is possible to change the default search circle radius as well as the shape and the size of the plotted points of that catalog.

In the **Options** column there are 3 letters:
**V**: to View the catalogue entries, in order to select or de-select specific points.
**S**: to save data of that specific catalog (for registered users)
**U**: to update catalog data, if changed.

For each catalog the user can change the default **colour** used in the displayed SED just by clicking on the coloured button next to the catalog Name.

**Box 1** The "Input Data" frame.

## 4.2 Plot Options

Plot settings can be changed from this tab (Figure 3). After clicking the **Load Data** button, your SED will be plotted in a log log scale, where $\nu f_\nu$ are in units of erg cm$^{-2}$ s$^{-1}$ and the frequencies in Hz. X and Y axis ranges can be changed and a fine tuning can be performed with the help of the red arrows on the left and right of the "Plot Axis Interval" box.

It is possible to zoom in (and after zoom out) on a particular SED area where data are crowded by placing the mouse near a data point and then, when the logo appears, by clicking with the mouse. With the same method you can also delete a data point. The point will be deleted only from the plot but not from the catalog, so it can be re-loaded if required. The displayed SED can be saved as an ASCII file in QDP format by clicking on "Export SED". The button "Save Image" enables to save the plotted SED in Portable Network Graphic (png) format.

The plot legend, below the main SED plot, and the plotted upper limits, can be erased if desired with an easy click (Figure 3).

There is also the possibility to reset all the catalog colors to one color only that can be selected by the user through the "Reset Catalog color" button. Also the background color can be changed (Figure 3).

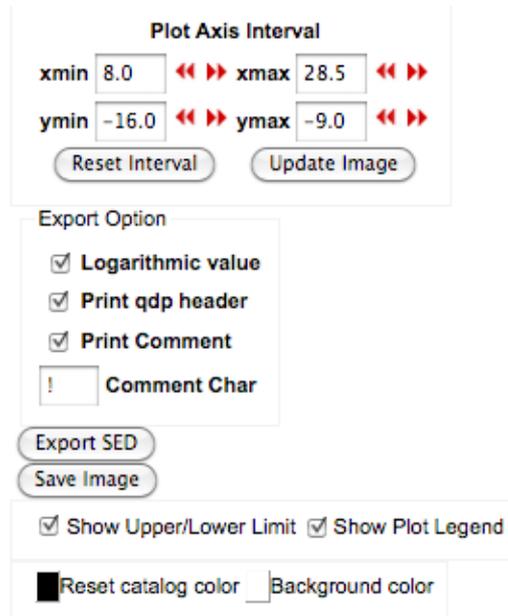

**Figure 3** The "Plot Options" functionality. A QDP file can be saved with "Export SED" while "Save Image" save the plot in 'png' format.

## 4.3 Models

This tab enables to define an emission model that can be compared with the displayed SED. Two types of Synchrotron Self Compton emission model are available:

1. **SSC Analytical** provides an analytic solution of the emission model. This solution, obtained through a number of approximations, has been developed by A. Pellizzoni and M. Perri. This model is based on a broken power law electrons distribution;
2. **SSC Numerical** provides a numerical approach to the model. It allows the user to choose among different electron distributions (e.g. power law, broken power law, log parabola, etc.). The user can select the model accuracy level and whether to consider or not the Synchrotron self-absorption component. Publications using the code should cite the following paper: A. Tramacere et al 2009, A&A, 501, 879.

***Both Analytical and Numerical SSC models require a non zero redshift***. The user can plot also both models together over the data by clicking on "Create new" button. In this case, a legend with the two models and the selected parameters (displayed by clicking the "+" symbol) will appear at the top of the frame (see Box 2).

To sum SSC components, and in general to sum existing single models in order to build a comprehensive model, click on "Create Model Grouping ": a new box in the top frame will be opened, allowing the user to select models for the model grouping.

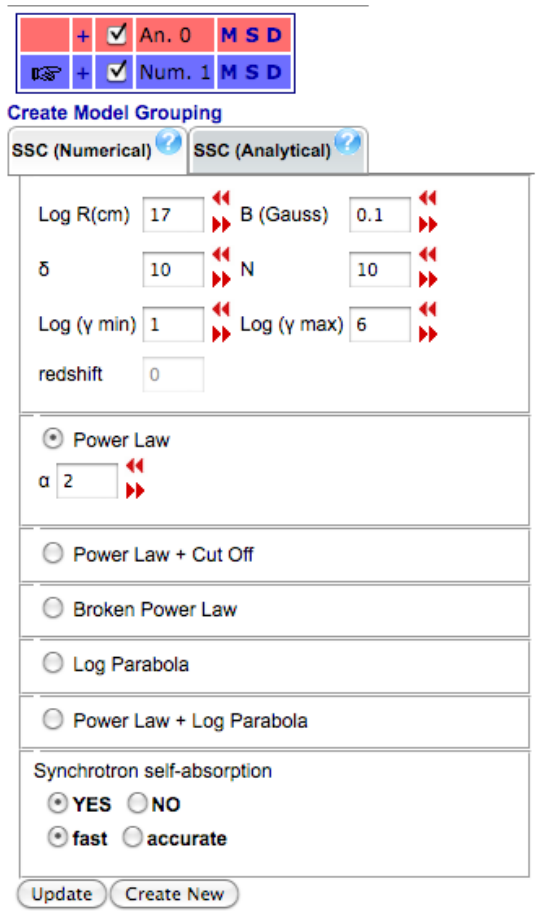

Here there is an example of an SSC Numerical model (Tramacere et al. 2009) frame. After parameters definition, the SED model will be displayed in the plot by clicking on "Update" button.

"Create New" enables to plot another model. Uploading new model makes a box with the same colour of the plotted model to appear above the parameter frame. The first column of the box indicates which model is active in the window below. The second one provides all the parameters setting, the third column enable to deselect one specific model, the fourth column indicates the name of the model. In the last column a number of options are provided: delete (D), save (S) and modify (M) the respective model.

"Create Model Grouping" enables to sum existing single models.

**Box 2** Models frame where the parameters of the Numerical SSC model are displayed.

## 4.4 Functions

The tab labelled "Functions" enables to fit a third degree polynomial function to the displayed data in a defined frequency range. The main goal of this functionality is to enable the user to estimate at best any peak flux value in the displayed SED: this value is indeed quoted as "Max value" at the bottom of the frame as soon as the fit has been performed (Figure 4).

As for the SSC models (see Section 4.3), more than one polynomial function can be plotted. In this case, clicking on "Draw new line", a legend with the two models will appear at the top of the frame.

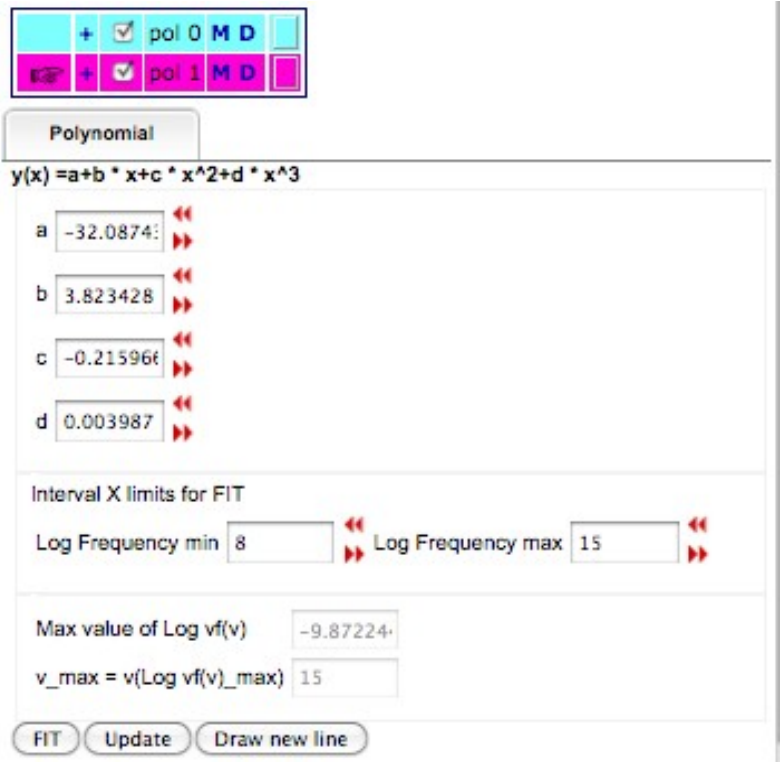

**Figure 4** The "Functions" frame for two polynomial functions uploaded with different colors. The frequency interval defines the fit range while the peak flux value found in that frequency range is quoted for the last fit as "Max value", together with the frequency at the maximum flux. Colors of each plotted line can be changed through the last column of the summary table at the top of the frame. This table appears when more than one function is used.

## 4.5 Templates

This functionality provides some useful templates to be compared with the displayed SED (Figure 5). A composite QSO optical spectrum from SDSS (D.E. Vanden Berk, 2001, ApJ, 122,549) plus Standard X-ray emission is available: this template requires a radio to optical flux ratio parameter to be defined.
In addition, a giant elliptical galaxy template is also available for comparison.

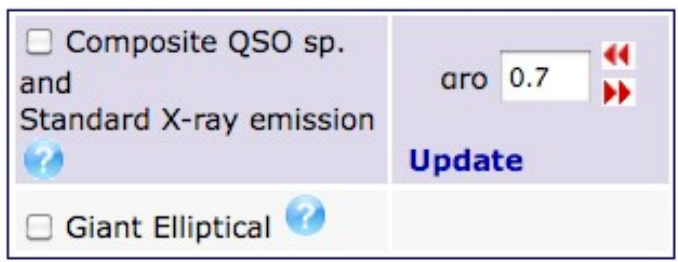

**Figure 5** The "Templates" frame

## 4.6 Instrumental sensitivity

This functionality enables to compare data with the sensitivity curves of several popular instruments, computed assuming exposure times equal to the deepest observations foreseen for each instrument. For gamma-ray (>100 MeV) detectors (AGILE, Fermi, EGRET) the limiting sensitivity corresponds to high galactic latitude and 1 year of observation. Cherenkov telescope sensitivies are for 50 hours of observation. For NuSTAR, an exposure of $10^6$ s was assumed. INTEGRAL ISGRI and PICsIT curves refer to $10^6$ s of exposure (3 σ). Planck sensitivity curves refer to point source 5-sigma sensitivity thresholds estimated from the [Planck Bluebook (Planck Collaboration 2005)](#) where background confusion is not included.

By selecting each instrument, a sensitivity curve is displayed in the SED plot (Figure 6). We warn the user that an updating work of some instrument sensitivity curve is still in progress.

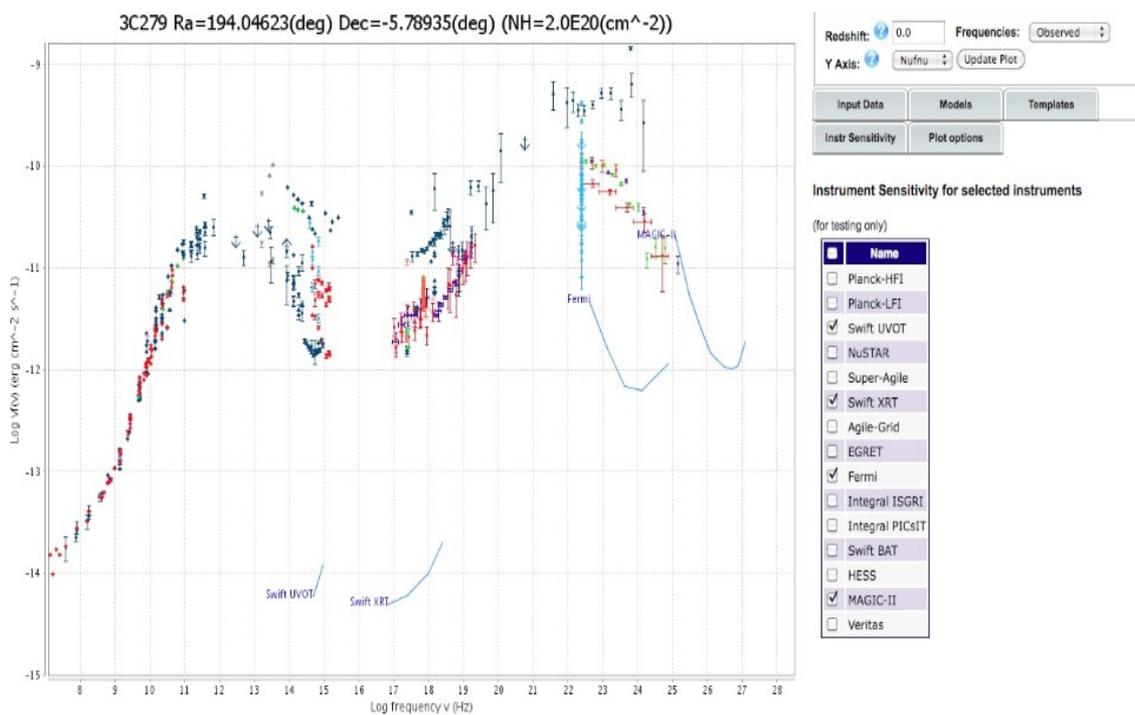

**Figure 6** The Instrumental Sensitivity section enables to compare the displayed SED with the instrumental sensitivity curves for several popular instruments.

## 4.7 SEDS

This tab enables the user to list all the user's saved SEDs previously built within a circular region centered on the current SED position. The default radius is 1 arcmin, but the user can change it at any time.

# 5. User Data Area

The **User Data** action button on the top right of the "ASDC SED Builder" page, redirects the user to the User Data Area where it is possible to upload proprietary (or any preferred) data.

**Figure 7** The "User Data" area.

The user can upload single source new data as well as multiple sources.

In the User Data area (Figure 7), the user can upload new data using a simple ASCII file where frequency in Hz units, flux and optionally flux errors in erg cm$^{-2}$ s$^{-1}$ units should be quoted in three separated columns: this format is specified in the quick help text box that appears moving the cursor on the question mark symbol. If user's new data refer to multiple sources, the first and second columns of the ASCII file should contain the RA and Dec of each source in degrees units.
Fluxes can be given also in logarithmic scale and flux errors can be ignored, if the respective boxes are ticked. Upper or Lower limits are also managed simultaneously with fluxes, by tagging the flux value with the label "UL" or "LL", respectively: if flux errors option has been selected, upper or lower limits should have a specified (e.g. null) error.

After uploading, the new data will be listed in a table at the bottom of the page (see Figure 8). In the "Options" column, the letter **D** and **E** stand for Delete and Edit. To erase a set of data, just to click on D letter. The Edit option allows the user to set any preferences on the colours and name of the catalog and enables also to share the data with other working groups and single users (see Section 5.2 for further details).

To see the new data in the SED plot, the user should go back to "Current SED", click on "Save" button and "Load" data again.

The user should remember that this set of data will be shown as "User Catalogs" in the "Input Data" section for each SED built with sky coordinates within 1 arcmin from the input coordinates and it will be available to be loaded on the SED if the input coordinates are within the catalogs "Default Search Radius".

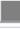

**Figure 8** User Data Table

## 5.1 Own SEDs

Every SED the user has built and saved is listed in this area. The user can access directly to saved SEDs from the starting page by clicking on "Own SEDs" on the top right (or from the query page, by clicking first on "SED" on the top right and then on "Own SEDs"). From this page, the user can Edit, Delete or View each SED from the list. 'Edit' allows to change the name and to share the SED with other collaborators (who need to be registered to the "ASDC SED Builder" tool).
Remember to click on 'Update SED' button to make effective your changes.

To go back to SED area, click on "current SED" link at the top of the page.

## 5.2 Sharing SED and User Data

To share a SED the user needs to, both:

- share the User Data Catalogs which contains data belonging to that source
- share the SED itself

This can be done by clicking on "Edit" on your own SED in the User Data area, selecting the desired group/users in the "Groups Not Allowed" list and clicking on "Join" button to move them to "Group Allowed list". To remove the sharing just select the User to be removed and click on "Leave" (Figure 9). User Data and SED are modifiable only by the creator of them, not by the Groups Allowed which benefit from a sharing.

**Catalog : pks2155-304_archive (Catalog ID = 105)**

| Groups not Allowed | Groups Allowed |
|---|---|
| Join | Leave |
| asdc<br>ErrorCircleAdmin<br>errorcircleanonymous<br>SedAdmin<br>sedanonymous<br>UserManagerAdmin | Fermi<br>perri<br>verrecchia |

**Figure 9** An example of User Data editing

# 6. Threads

In this section a number of threads are provided.

## 6.1 Quick comparison of new data with archival ones

This example may be useful when new, fresh data need to be compared with archival ones.
For this example, the user should be registered and logged in.

1) From the starting page of the "ASDC SED Builder" tool (http://tools.asdc.asi.it/SED/), click on the indicated link to start to build a SED.
2) In the query page, provide the name of the source (or its coordinates), and then click on "Search by name" (or "Search by coordinates"). An output table will appear at the bottom of the page
3) The last column of the table, named "Tools", provides two links:
    a. "Data Explorer" redirects the user to a new tab where the "Data Explorer" ASDC Tool is loaded on the input coordinates.
    b. "Create SED" redirects the user to the main "SED Builder" page, that will be opened in a new tab.
   Click on "Create SED" to continue.

4) Click on the "Load Data" action button. This will plot the SED built with all



\end{document}